\RequirePackage{fix-cm}
\documentclass[smallextended]{svjour3}       
\smartqed  
\usepackage{graphicx}
\usepackage{mathtools}
\usepackage{float}
\usepackage{multirow}
\usepackage{lineno,hyperref}
\usepackage{amsmath,amssymb,amstext,tcolorbox}
\usepackage{graphicx}
\usepackage[utf8]{inputenc}
\usepackage{array}
\newcolumntype{P}[1]{>{\centering\arraybackslash}p{#1}}

\begin{document}

\title{Predicting user demographics based on interest analysis
}


\author{Reza Shafiloo         \and
        Marjan Kaedi \and
        Ali Pourmiri
}


\institute{Reza shafiloo \at
              Faculty of Computer Engineering, University of Isfahan, Azadi Sq., Hezarjarib St., Isfahan, Iran \\
              \email{rezashafilou@eng.ui.ac.ir}           
           \and
           Marjan Kaedi \at
              Faculty of Computer Engineering, University of Isfahan, Azadi Sq., Hezarjarib St., Isfahan, Iran
              \email{kaedi@eng.ui.ac.ir}
              \and
              Ali Pourmiri \at
              Macquarie University, Sydney, Australia
              \email{ali.pourmiri@mq.edu.aur}
}

\date{Received: date / Accepted: date}

\maketitle

\begin{abstract}
These days, due to the increasing amount of information generated on the web, most web service providers try to personalize their services. Users also  interact with web-based systems in multiple ways and state their interests  and preferences by rating the provided items.
In this paper, we propose  a framework to predict users' demographic  based on ratings registered by users in a  system. To the best of our knowledge, this is the first time that the  item ratings are employed for users' demographic prediction problem, which has extensively been studied  in recommendation systems and service personalization.
We apply the framework to Movielens dataset's ratings  and predict users' age and gender.    The experimental results show that using all ratings registered by users improves the prediction accuracy by at least 16\% compared with previously studied models. Moreover, by classifying the items as popular and unpopular,  
we eliminate ratings belong to 95\% of items and still  reach an acceptable level of accuracy. This  significantly reduces  update cost in a time-varying environment. Besides this classification, we propose other methods to reduce data volume while keeping the predictions accurate.

\end{abstract}
	\section{Introduction}
With the development of technology, people tend to use internet-based systems – such as e-learning systems, e-commerce, and social networks – more often. The quantity of content produced in these systems, therefore, is increasing dramatically. In these systems, the services are personalized according to users' previous preferences. Consequently, users gain a better experience by interacting with a personalized system \cite{wang2016multi}. It seems necessary to obtain information about users to personalize services. Users' age and gender are the  most useful information for personalizing services. Classifying users based on their age and gender can be valuable in providing content closer to their interests \cite{hu2007demographic}. However, most users of these systems are not interested in displaying their personal information – including age and gender. It is necessary, then, to identify and estimate the user's demographic based on his/her interactions \cite{guimaraes2017age}. Users demographic information have been identified  based on  posts they publish and sentiment analysis on the posts \cite{pandya2020use}, and  web browsing patterns \cite{hu2007demographic}. Also, obtaining such information about users will enable recommender systems to provide more relevant contents to their users with higher accuracy. 

In this work we focus on identifying   
users' age and gender based on their interests in the provided contents  which  has not been identified yet – at least to the best of our knowledge.
\subsection{Motivation and paper contribution}
\begin{figure}[!t]
		\centering
		\includegraphics[scale=0.6]{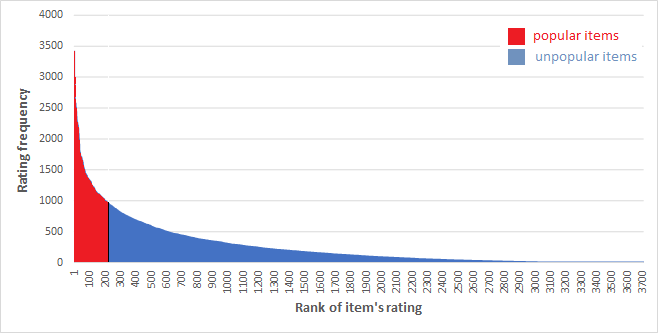}
		\caption{ The figure illustrates the  rating frequency  for each movie (i.e., from 1 to 3700). As one can see  the first 5\% of movies  get rated at least 1000 times that is called the popular items/movies. The rate frequency of each movie is the number times that a movie gets rated by users. }\label{Fig:longtail}
	
\end{figure}

To predict the demographic information, previous works
\cite{garcia2020trend,nguyen2013old,morgan2017predicting,kalimeri2019predicting,kim2019predicting,zhong2013user,dong2014inferring,al2019predicting,li2019improving,bin2017identifying,huang2017age,malmi2016you}
employ all data generated by users. On the other hand,  
information extraction based on  users interests seems to be challenging as data is \emph{massive} and \emph{time-varying}. For instance in recommender systems, one can observe that 
the  data volume gathered from user preferences increases over time, and  
newly registered interests by an user may not be consistent with the previous user interests.
Therefore,  the time of rebuilding new models for rising amount of data increase over time and the general  data reduction techniques such as \emph{prototype selection} (e.g., see \cite{garcia2012prototype})  may not provide a robust solution.
In this work we introduce a new framework for user information prediction in a  massive and time-varying environment.
Our framework is based on a phenomenon that has been observed in many human-intarction systems for which   users react to a subset of  contents, so-called the \emph{popular items}, more frequently.
This classification of popular and unpopular items can also be found in newspaper-related databases \cite{diez2019optimizing}, e-commerce systems \cite{huang2019novel}, libraries \cite{valcarce2016item}, and social networks \cite{eirinaki2018recommender}. Moreover, it has been observed  that the item's popularity follows a power law distribution \cite{ahmadian2020social} which indicates that the  popular items are only  a small fraction of  all items.
Using the popular items have two advantages, first it only considers a relatively small portion of   the data, second our  experimental analysis shows that they are robust in time-varying datasets.
In order to show the efficiency of the  proposed frameworks,  we identify age and gender based on the user's interest in movies in the Movielens dataset and movie features in the IMDB database. 
Similar to aforementioned examples, our data set can also be divided to popular and unpopular items.
Figure \ref{Fig:longtail} is a histogram of the ratings for movies  in Movielens dataset with one million ratings  that users have recorded  \cite{hamedani2019recommending}.  Notice that 
each rate recorded for a  movie reflects  a  user's opinion about the movie.
According to this Figure, almost 30\% of ratings are related to popular movies that are only  5\% of all movies. 
\begin{figure}
	\centering
	\includegraphics[scale=0.6]{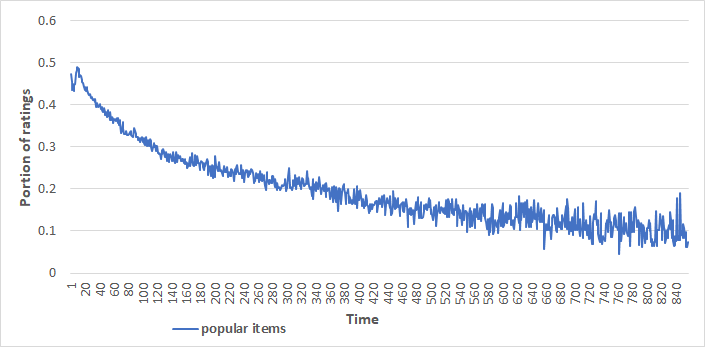}
	\caption{The figure illustrates the portion of rating registered for popular movies in terms of  chronological ordering of ratings.
		For instance, it shows that around 50\% of the first ratings of all users hs been registered for popular movies and this number decreases over time.
	}\label{portionLTSHtimeOrder}\label{Fig:portion}
\end{figure}
Moreover, Figure \ref{Fig:portion} shows  that users tend to rate popular movies in their early ratings. After a while, users register ratings mostly for unpopular movies and rate for popular movies rarely. Moreover, due to the phenomenon, called  "the rich get richer", popular items will become more famous \cite{taeuscher2019uncertainty}. So, popular items usually remain famous and it is less common for popular items to become unpopular. Therefore, using  the popular items for user modeling reduces the regular updates of  trained models and hence it is cost-effective. 

In this paper, we  identify user's demographic by  analyzing  ratings recorded for all movies and get an accuracy of 65\% which is as precise as  previous results \cite{li2019improving,al2019predicting} and we get an accuracy of 80\% for gender prediction. Moreover, by restricting ratings to those recorded for popular movies (which include 5\% of all movies), we acquire an accuracy of 60\% for age prediction. Furthermore, we determine user's gender by using ratings that are registered for popular movies and the accuracy   is improved almost 8\% compared with a recent  gender prediction  \cite{garcia2020trend}. For a more detailed discussion regarding the metrics we refer the reader to Subsection \ref{sub:disc}.
\subsection{Related works}\label{sec:lit}
Various methods have been used in previous literature to identify users' demographic information. Two of the most important demographic information to be considered is users' age and gender. Previous studies have considered different approaches to identify user's age and gender based on different input data.

One of the most common platforms for users' demographic identification is social networks, and particularly, Twitter. Nguyen et al. \cite{nguyen2013old} identify Twitter users' age using textual information posted by users. Morgan-Lopez et al. \cite{morgan2017predicting} perform user's age modeling using tweets and metadata such as the number of friends and the number of followers. The combination of these two types of data has achieved the best results in their study. Guimaraes et al. \cite{guimaraes2017age} in addition to textual information obtained from users' tweets, predict users' age based on other information such as Twitter followers' network information, the number of tweets, and also user's gender. Li et al. \cite{li2019improving} design an attention-based neural network to identify features such as age, gender, and geographic location by extracting the meanings of texts published by users and social network information. Garcia et al. \cite{garcia2020trend} developed two models for determining a user's age and gender. The first model is based on textual information posted by users on Twitter and the other one is based on the category of photos that users have posted on Pinterest. Pandya et al. \cite{pandya2020use} use  the contents of the URLs and hashtags that are used in the tweets to classify users' age .

In addition to social networks, web service providers have decided to personalize their services. Hu et al. \cite{hu2007demographic} identify users' age and gender by data obtained from users' clicks on web pages, page categories, and words used on the page. Kalimeri et al. \cite{kalimeri2019predicting} obtain information about the pages browsed on the user's personal computer and mobile phone, and the programs used on the user's mobile phone, then, perform age and gender modeling by demographic data that users have voluntarily reported in their study. Kim et al. \cite{kim2019predicting} determine user's demographic information by the textual content of the websites visited by the users and the information that users have registered in the database.

Some other studies use mobile data such as calls and SMS to identify users' age. Zhong et al. \cite{zhong2013user} introduce a five-part probabilistic framework and demographic information modeling (including age and gender) based on textual features, multitasking modeling, and cost-sensitive modeling. Dong et al. \cite{dong2014inferring} present a new probabilistic model based on the characteristics of mobile users' communications (such as calls and SMS) and their behavior in the mobile network. Al-Zuabi et al. \cite{al2019predicting} use type of services that is running on mobile and social behavior features to identify user's age and gender. 

Also, there are other categories of data which are used for demographic prediction. Tareaf et al. \cite{bin2017identifying} predict users' demographic based on the common words that they use and 82 writing styles. Huang et al. \cite{huang2017age} use the deep learning network to identifying user age based on their images. Malmi et al. \cite{malmi2016you} determine Android users' age and gender based on the Android applications that utilized by user.

\subsection{Paper structure} 
In Section \ref{sec:data}, we explain the proposed methods for the age and gender identification based on users' ratings. In Section \ref{sec:res}, we evaluate the proposed methods. Finally, a summarization of the paper is presented in Section \ref{sec:dis}.

\section{The proposed method to identify the user's age and gender}\label{sec:data}
In this section, we first introduce the dataset used in this study, and then  we propose three methods to  apply  users’ rating in age and gender prediction. Finally, we  present machine learning algorithms for the demographic identification problem.
\subsection{Data description} \label{sec:dataDes}
Movielens dataset \cite{noauthor_movielens_2013} has been  used in several studies and products. We use a version of Movielens dataset, called ML 1M, that includes one million ratings  and contains users' demographic information such as  age, gender, occupation, and zip-code. ML 1M has 6040 users, 3883 movies, and ratings that have been registered for three years (i.e., from 2000 to 2003). In what follows we explain how data are represented in this dataset. Throughout this paper we assume that movies and users are represented by unique positive numbers  $i$   and $u$, respectively. ML 1M dataset contains the following tuple
$T=(u,i,r,t)$
which indicates  user $u$ registers rating $1\le r\le 5$ for movie $i$ at time $t$. Additionally, each movie is labeled by  eighteen different genres, namely action, adventure, animation, children's, comedy, crime, documentary, drama, 
fantasy, film-noir, horror, musical, mystery, romance, sci-fi, thriller, war, and western. Therefore, for every movie  $i$, one can define $18$-dimensional vector 
$(g_1^i,\ldots,g_{18}^i)$
where, for every $1\le j\le 18$, $g_j^i$ is set to be $1$ if movie $i$  is associated with genre $j$ and zero otherwise. Notice that a movie might be associated with several genres. Along with Movielens dataset, we apply   information  that is available in the IMDB database. In this study we apply two types of information provided by IMDB \cite{imdb}. 

The first is an age rating that is registered by Motion Picture Association Rating System \cite{motion}. Here, the age rating indicates the film suitability for different ages based on its content. Films get labeled by G, PG, PG-13, R, NC-17, and unrated. Therefore, for each movie $i$, one can  define $6$-dimensional vector  $(m_1^i,\ldots,m_6^i)$
where, for every $1\leq j\le 6$, $m_j^i$  is   $1$ if movie $i$  belongs to category $j$  and $0$ otherwise. For instance,  if movie $i$ gets labeled by G, then $m_1^i=1$.
The second useful information is the parental guide information from IMDB database that is  associated with each movie $i$ and denoted by $(p^i_1,\ldots,p_5^i )$. The vector measures particular contents of a movie including sex \& nudity, violence \& gore, profanity, alcohol \& drug \& smoking, and frightening \& intense scenes \cite{imdbp}. For every $1\le j\le 5$ , $p_j^i\in [0,3]$  is a score given to the movie based on the  $j$-th content. For instance, $p_1^i=0$ means that movie  $i$ has the  lowest  possible sex \& nudity content.
\subsection{Feature vectors}\label{sec:Feat}
Using the data described in the previous subsection, corresponding to each user $u$ we define  three feature vectors, namely $X(u)$, $Y(u)$, and $Z(u)$.
Recall that dataset contains a set of  tuples $T=(u,i,r,t)$ defined in the previous subsection.
Considering all tuples, for every user $u$, we define  $S_u$ to be the set of all movies rated by user $u$.
\paragraph{All movies feature vector}
Define $X(u)=(x^u_j)_{j=1}^{30}$ to be a $30$-dimensional vector which is defined as follows:
\begin{align}\label{fv}
x^u_j=
\begin{cases} 
	\frac{\sum_{i \in S_u} g^i_j}{|S_u|} & 1\le j\le 18, \\
	\frac{\sum_{i \in S_u} m^i_{j-18}}{|S_u|} & 19\le j\leq 24, \\
	\frac{\sum_{i \in S_u} p^i_{j-24}}{|S_u|} & 25\le j\leq 29,\\
\end{cases}
\end{align}

where $g^i_j$, $m^i_j$ and $p_j^i$  are corresponding to genres, age categories and contents of movie $i$, respectively, defined in the previous subsection. Also, $x^u_{30}$ is user age category.
\paragraph{$\alpha$-Popular movies  feature vector}  
Here, we consider  a popular subset of movies rated by user $u$ defined as follows:
Suppose that all movies are decreasingly  sorted base on their average rating score (i.e., averaging over  ratings given by all users). The first $\alpha$ fraction of the movies in the sorted list  is  called $\alpha$-popular, denoted by $P_\alpha$.
Considering (\ref{fv}), we define  $\alpha$-popular movies  feature vector, for which  $S_u$ is replaced by $S_u\cap P_\alpha$. The vector is denoted by $Y(u)=(y^u_j)_{j=1}^{30}$ and difined as follows:
\begin{align}\label{fv2}
	y^u_j= \begin{cases} 
		\frac{\sum_{i \in S_u\cap P_\alpha} g^i_j}{|S_u\cap P_\alpha|} & 1\le j\le 18, \\
		\frac{\sum_{i \in S_u\cap P_\alpha} m^i_{j-18}}{|S_u\cap P_\alpha|} & 19\le j\leq 24, \\
		\frac{\sum_{i \in S_u\cap P_\alpha} p^i_{j-24}}{|S_u\cap P_\alpha|} & 25\le j\leq 29,\\
	\end{cases}
\end{align}

\paragraph{Liked movies feature vector}
Let $r_u(i)$ denoted the rate given by user $u$ to movie $i$.
In order to find the user's favorite items, we calculate the average ratings per user denoted by $\mu_u$  as follows 
\begin{align*} 
	\mu_u=\frac{\sum_{i \in S_u} r_u(i)}{|S_u|} 
\end{align*}
Now, corresponding to each user $u$, we define a set of liked movies as follows:
\[
L_u=\{i\in S_u: r_u(i)\ge \mu_u\}.
\]
Clearly, we have that  $L_u\subseteq S_u$. Considering (\ref{fv}), we replace $S_u$ by $L_u$ and define  a liked movies feature vector denoted by $Z(u)$.
\subsection{Applying machine learning algorithms to data}
After data organization with the different methods mentioned in Section \ref{sec:Feat}, various machine learning algorithms apply to the data in order to extract models for predicting user's age and gender in a supervised learning process. We apply some of the best algorithms used in previous studies on our prepared vectors, namely $K$-Nearest Neighbor, Naïve Basie, Random Forest, Multilayer Perceptron, and XGBoost algorithms which are referred as KNN, NB, RF, MLP, and XGB in Section \ref{sec:res}. Their results are evaluated and compared in Section \ref{sec:res}.
\section{Results and discussion}\label{sec:res}
In this section, the three different methods presented in Section \ref{sec:Feat}, and the application of the mentioned machine learning functions are evaluated and compared. Other works consider different $\alpha$ in their works for popular movies \cite{sreepada2020mitigating,park2008long,hamedani2019recommending}. To obtain our result, we consider 5\%-popular movies to show ratings related to this amount of popular movies reduces over time and converges to zero.
\subsection{Evaluation metrics}\label{sec:evMet}
Various metrics are used to assess the accuracy of user's age and gender classification. In this study, we use the accuracy criterion for this purpose \cite{japkowicz2011evaluating}. In the following Equation we calculate the accuracy of classifier $f$ on test set $T$. Let $C(f)$ denotes an $l\times l$ matrix where $l$ is the number of classes in the dataset. Also, $c_{ij}(f)$ denotes the number of samples with actual class $i$ assigned to a class $j$ by classifier $f$.
\[Accuracy_f(T)=\frac{\sum^l_{i=1} c_{ii}(f)}{\sum^l_{i,j=1} c_{ij}(f)}, \]
where $\sum^l_{i=1} c_{ii}(f)$ denotes total number of samples correctly classified by classifier $f$, and $\sum^l_{i,j=1} c_{ij}(f)$ show total number of predictions are made by classifier $f$. Moreover, we use precision metric which is defined as follows. 
\begin{align*} 
	Precision_i(f)=\frac{c_{ii}}{\sum^l_{j=1} c_{ji}(f)},
\end{align*}
where $c_{ii}$ shows number of observations belong to actual class $i$ and correctly predicted  classified by classifier $f$, and $\sum^l_{j=1} c_{ji}(f)$ denotes total number of observations that predicted as class $i$. Next metric is recall that denotes
\begin{align*} 
	Recall_i(f)=\frac{c_{ii}}{\sum^l_{j=1} c_{ij}(f)} 
\end{align*}
which $\sum^l_{j=1} c_{ij}(f)$ shows all observations in actual class $i$.
Then, we use F measure as follow
\begin{align*} 
	F1-score_i(f)=\frac{2 \times Precision_i(f) \times Recall_i(f)}{Precision_i(f) + Recall_i(f)} 
\end{align*}
which $F1-score$ shows harmonic mean of precision and recall. To evaluate the results, in addition to accuracy, weighted precision and weighted F-measure are used as follow. 
\begin{align*} 
	Weighted-precision=\sum^l_{k=1} w_{k}Precision_k(f)
\end{align*}
\begin{align*} 
	Weighted-F1-score=\sum^l_{k=1} w_{k}F1-score_k(f)
\end{align*}
$w_{k}=\frac{|k|}{|T|}$ shows the weight of each class in test set which $|T|$ shows test set size, and $|k|$ number of samples blong to class $k$ in test set $T$.

\subsection{Results obtained for age and gender in Movielens dataset}
In this section, the results of algorithms which applied on age and gender categories registered in the Movielens dataset (described in Section \ref{sec:dataDes}), are presented in separate tables based on the criteria introduced in Section \ref{sec:evMet}.
\begin{table}[h!]
	\centering
	\caption{Accuracy of machine learning algorithms used to predict user's age and gender for three different methods}\label{Table:Acc7}
	\begin{tabular}{ | P{13em} | P{0.7cm}| P{0.7cm}| P{0.7cm} | P{0.7cm} | P{0.7cm} | P{0.7cm} | } 
		\hline
		Methods&& KNN & NB & RF & MLP& XGB  \\ 
		\hline
		\multirow{2}{13em}{All Items: 1000209 ratings 3883 items} & \multicolumn{1}{c|}{Age} & %
		\multicolumn{1}{c|}{0.32} & \multicolumn{1}{c|}{0.31}&\multicolumn{1}{c|}{0.37}&\multicolumn{1}{c|}{0.41}&%
		\multicolumn{1}{c|}{0.38}\\
		\cline{2-7}
		&  \multicolumn{1}{c|}{Gender} & %
		\multicolumn{1}{c|}{0.75} & \multicolumn{1}{c|}{0.73}&\multicolumn{1}{c|}{0.77}&\multicolumn{1}{c|}{0.80}&%
		\multicolumn{1}{c|}{0.77} \\
		\hline
		\multirow{2}{14em}{Popular items: 299254 ratings 202 items} & \multicolumn{1}{c|}{Age} & %
		\multicolumn{1}{c|}{0.31} & \multicolumn{1}{c|}{0.32}&\multicolumn{1}{c|}{0.35}&\multicolumn{1}{c|}{0.36}&%
		\multicolumn{1}{c|}{0.37}\\
		\cline{2-7}
		&  \multicolumn{1}{c|}{Gender} & %
		\multicolumn{1}{c|}{0.72} & \multicolumn{1}{c|}{0.70}&\multicolumn{1}{c|}{0.75}&\multicolumn{1}{c|}{0.75}&%
		\multicolumn{1}{c|}{0.73} \\
		\hline
		\multirow{2}{13em}{Liked items: 545279 ratings 3883 items } & \multicolumn{1}{c|}{Age} & %
		\multicolumn{1}{c|}{0.29} & \multicolumn{1}{c|}{0.32}&\multicolumn{1}{c|}{0.34}&\multicolumn{1}{c|}{0.35}&%
		\multicolumn{1}{c|}{0.31}\\
		\cline{2-7}
		&  \multicolumn{1}{c|}{Gender} & %
		\multicolumn{1}{c|}{0.74} & \multicolumn{1}{c|}{0.72}&\multicolumn{1}{c|}{0.77}&\multicolumn{1}{c|}{0.79}&%
		\multicolumn{1}{c|}{0.77} \\
		\hline
	\end{tabular}
	
\end{table}
\begin{table}[h!]
	\centering
	\caption{Weighted precision of Machine learning algorithms used to predict user's age and gender for three different methods}\label{Table:per7}
	\begin{tabular}{ | P{13em} | P{0.7cm}| P{0.7cm}| P{0.7cm} | P{0.7cm} | P{0.7cm} | P{0.7cm} | } 
		\hline
		Methods&& KNN & NB & RF & MLP& XGB  \\ 
		\hline
		\multirow{2}{13em}{All Items: 1000209 ratings 3883 items} & \multicolumn{1}{c|}{Age} & %
		\multicolumn{1}{c|}{0.28} & \multicolumn{1}{c|}{0.31}&\multicolumn{1}{c|}{0.34}&\multicolumn{1}{c|}{0.38}&%
		\multicolumn{1}{c|}{0.36}\\
		\cline{2-7}
		&  \multicolumn{1}{c|}{Gender} & %
		\multicolumn{1}{c|}{0.73} & \multicolumn{1}{c|}{0.73}&\multicolumn{1}{c|}{0.75}&\multicolumn{1}{c|}{0.79}&%
		\multicolumn{1}{c|}{0.75} \\
		\hline
		\multirow{2}{14em}{Popular items: 299254 ratings 202 items} & \multicolumn{1}{c|}{Age} & %
		\multicolumn{1}{c|}{0.28} & \multicolumn{1}{c|}{0.28}&\multicolumn{1}{c|}{0.3}&\multicolumn{1}{c|}{0.28}&%
		\multicolumn{1}{c|}{0.33}\\
		\cline{2-7}
		&  \multicolumn{1}{c|}{Gender} & %
		\multicolumn{1}{c|}{0.68} & \multicolumn{1}{c|}{0.69}&\multicolumn{1}{c|}{0.73}&\multicolumn{1}{c|}{0.73}&%
		\multicolumn{1}{c|}{0.71} \\
		\hline
		\multirow{2}{13em}{Liked items: 545279 ratings 3883 items } & \multicolumn{1}{c|}{Age} & %
		\multicolumn{1}{c|}{0.24} & \multicolumn{1}{c|}{0.29}&\multicolumn{1}{c|}{0.29}&\multicolumn{1}{c|}{0.29}&%
		\multicolumn{1}{c|}{0.26}\\
		\cline{2-7}
		&  \multicolumn{1}{c|}{Gender} & %
		\multicolumn{1}{c|}{0.72} & \multicolumn{1}{c|}{0.73}&\multicolumn{1}{c|}{0.76}&\multicolumn{1}{c|}{0.78}&%
		\multicolumn{1}{c|}{0.76} \\
		\hline
	\end{tabular}
\end{table}
\begin{table}[H]
	\centering
	\caption{Weighted average F-measure of Machine learning algorithms used to predict user's age and gender for three different methods}\label{Table:F1-7}
	\begin{tabular}{ | P{13em} | P{0.7cm}| P{0.7cm}| P{0.7cm} | P{0.7cm} | P{0.7cm} | P{0.7cm} | } 
		\hline
		Methods&& KNN & NB & RF & MLP& XGB  \\ 
		\hline
		\multirow{2}{13em}{All Items: 1000209 ratings 3883 items} & \multicolumn{1}{c|}{Age} & %
		\multicolumn{1}{c|}{0.29} & \multicolumn{1}{c|}{0.29}&\multicolumn{1}{c|}{0.32}&\multicolumn{1}{c|}{0.35}&%
		\multicolumn{1}{c|}{0.35}\\
		\cline{2-7}
		&  \multicolumn{1}{c|}{Gender} & %
		\multicolumn{1}{c|}{0.72} & \multicolumn{1}{c|}{0.73}&\multicolumn{1}{c|}{0.75}&\multicolumn{1}{c|}{0.78}&%
		\multicolumn{1}{c|}{0.75} \\
		\hline
		\multirow{2}{14em}{Popular items: 299254 ratings 202 items} & \multicolumn{1}{c|}{Age} & %
		\multicolumn{1}{c|}{0.29} & \multicolumn{1}{c|}{0.27}&\multicolumn{1}{c|}{0.28}&\multicolumn{1}{c|}{0.27}&%
		\multicolumn{1}{c|}{0.33}\\
		\cline{2-7}
		&  \multicolumn{1}{c|}{Gender} & %
		\multicolumn{1}{c|}{0.68} & \multicolumn{1}{c|}{0.69}&\multicolumn{1}{c|}{0.72}&\multicolumn{1}{c|}{0.72}&%
		\multicolumn{1}{c|}{0.71} \\
		\hline
		\multirow{2}{13em}{Liked items: 545279 ratings 3883 items } & \multicolumn{1}{c|}{Age} & %
		\multicolumn{1}{c|}{0.25} & \multicolumn{1}{c|}{0.29}&\multicolumn{1}{c|}{0.26}&\multicolumn{1}{c|}{0.29}&%
		\multicolumn{1}{c|}{0.26}\\
		\cline{2-7}
		&  \multicolumn{1}{c|}{Gender} & %
		\multicolumn{1}{c|}{0.72} & \multicolumn{1}{c|}{0.73}&\multicolumn{1}{c|}{0.75}&\multicolumn{1}{c|}{0.78}&%
		\multicolumn{1}{c|}{0.76} \\
		\hline
	\end{tabular}
	
\end{table}
\subsection{Reducing age classes}
In this section, we reduce the number of age categories from 7 to 3, to compare our work with the age prediction methods presented in previous literature. New age categories are shown in Table \ref{Table:RedAge}.
	\begin{table}[H]
	\centering
	\caption{New age categories for Movielens Dataset }\label{Table:RedAge}
	\begin{tabular}{ | P{15em} | P{5cm}|  } 
		\hline
		Age label in Movielens dataset& New label  \\ 
		\hline
		1,18,25& young  \\ 
		\hline
		35,45 & adult  \\ 
		\hline
		50,56 & old  \\ 
		\hline
	\end{tabular}
	
\end{table}
In the following, the results obtained from our three different methods with new defined age categories are summarized in Tables \ref{Table:Acc3}, \ref{Table:per3}, and \ref{Table:F1-3}.
\begin{table}[h!]
	\centering
	\caption{Accuracy of machine learning algorithms used to predict user age based on new age categories for three different methods}\label{Table:Acc3}
	\begin{tabular}{ | P{7em} | P{0.7cm}| P{0.7cm} | P{0.7cm} | P{0.7cm} | P{0.7cm} | } 
		\hline
		Methods& KNN & NB & RF & MLP& XGB  \\ 
		\hline
		All Items 1000209 ratings 3883 items& 0.59 & 0.58 & 0.62 & 0.65 & 0.63  \\ 
		\hline
		Popular items 299254 ratings 202 items & 0.53 & 0.56 & 0.59 & 0.6 & 0.58  \\ 
		\hline
		Liked items 545279 ratings 3883 items & 0.54 & 0.55 & 0.57 & 0.56 & 0.58  \\ 
		\hline
	\end{tabular}
	
\end{table}
\begin{table}[h!]
	\centering
	\caption{Weighted precision of Machine learning algorithms used to predict user age  based on new age categories for three different methods}\label{Table:per3}
	\begin{tabular}{ | P{7em} | P{0.7cm}| P{0.7cm} | P{0.7cm} | P{0.7cm} | P{0.7cm} | } 
		\hline
		Methods& KNN & NB & RF & MLP& XGB  \\ 
		\hline
		All Items 1000209 ratings 3883 items& 0.56 & 0.53 & 0.59 & 0.61 & 0.61  \\ 
		\hline
		Popular items 299254 ratings 202 items & 0.48 & 0.52 & 0.55 & 0.53 & 0.54  \\ 
		\hline
		Liked items 545279 ratings 3883 items & 0.49 & 0.52 & 0.5 & 0.5 & 0.46  \\ 
		\hline
	\end{tabular}
	
\end{table}
\begin{table}[h!]
	\centering
	\caption{Weighted average F-measure of Machine learning algorithms used to predict user age  based on new age categories for three different methods}\label{Table:F1-3}
	\begin{tabular}{ | P{7em} | P{0.7cm}| P{0.7cm} | P{0.7cm} | P{0.7cm} | P{0.7cm} | } 
		\hline
		Methods& KNN & NB & RF & MLP& XGB  \\ 
		\hline
		All Items 1000209 ratings 3883 items& 0.56 & 0.53 & 0.58 & 0.61 & 0.61  \\ 
		\hline
		Popular items 299254 ratings 202 items & 0.5 & 0.52 & 0.53 & 0.54 & 0.55  \\ 
		\hline
		Liked items 545279 ratings 3883 items & 0.51 & 0.53 & 0.5 & 0.51 & 0.47  \\ 
		\hline
	\end{tabular}
	
\end{table}

\subsection{Discussion} \label{sub:disc}
In section \ref{sec:Feat} we introduce three different method to reduce data that is using for machine learning's input vector. In first method, we use all ratings, in the second one, we use ratings of $\alpha$-popular movies, and in the last second one we use ratings of liked movies.

According to the accuracy evaluation, the best performance for the seven age categories registered in the Movielens dataset is obtained by the MLP algorithm in which all of the data is preserved (first method). The classification accuracy in the other methods – i.e., the second and third methods – has decreased compared to the first method; but the accuracy reduction for the best algorithm (MLP) has been reduced to at least 0.35\%, which is only 6\% less than the best result for this algorithm. Therefore, a big part of the data is eliminated by considering the user's favorite movies or the ratings related to the popular items, while, the accuracy does not decrease significantly. Also, by considering popular items' ratings, we use information of only 5\% of all movies to predict users' age.

In the next step, to compare our work with other studies, we reduce the number of age classes to three. Yumeng Li et al. \cite{li2019improving}, Ibrahim Mousa Al‑Zuabi et al. \cite{al2019predicting} get an accuracy of 0.65 for the three age groups, and Roberto Garcia-Guzman et al. \cite{garcia2020trend} has an accuracy of 0.67 for the same number of age groups. In our study, by reducing the age groups from 7 to 3 categories, the accuracy of the MLP algorithm in the first method – i.e., using popular items ratings – has achieved an accuracy of 0.65\%. Besides, in the second method, the accuracy of the MLP algorithm has reached 0.6, in which only 30\% of the total ratings has been used, and only a five percentage accuracy reduction has been faced.

\begin{figure}[H]
	\centering
	\includegraphics[scale=0.6]{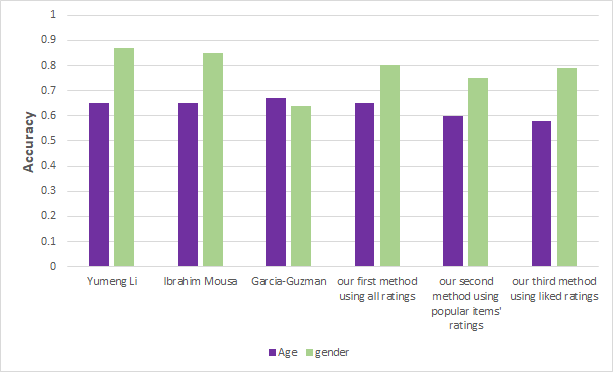}
	\caption{Comparing best result of Yumeng Li \cite{li2019improving}, Ibrahim Mousa Al ‑ Zuabi \cite{al2019predicting}, and Garcia-Guzman \cite{garcia2020trend} versus our proposed methods}\label{Fig:comparing}
\end{figure}

Also, we use our three methods for gender classification and get the best result from the first method with an accuracy of 0.8. Yumeng Li et al. \cite{li2019improving}, acquires an accuracy of 0.87, and Ibrahim Mousa Al‑Zuabi et al. \cite{al2019predicting} get an accuracy of 0.85 for gender prediction. The accuracy of gender identification for Roberto Garcia-Guzman et al. \cite{garcia2020trend} is 0.64, which is less than the accuracy of our second method – using popular items – with an accuracy of 0.75. This result indicates that using ratings of popular items can reach a desirable level of accuracy. We illustrate our results comparison on Figure \ref{Fig:comparing}.\\
In Figure \ref{Fig:portion}, we show users tend to register ratings for popular items in their early ratings. After a while, users rate unpopular movies more than popular ones. Therefore, $y^u_j$ need to be updated in the early ratings for user $u$, and after a while, it does not need to update. However, $x^u_j$ should be updated when user $u$ registers new ratings, and $z^u_j$ should be updated when user $u$ likes an item. Therefore, we can build prediction models based on users' popular ratings with acceptable accuracy that does not need regular updating during the time, and in the users' last ratings, it nearly does not need to update. 
\section{Conclusion and future study}\label{sec:dis}
Various studies have shown that users with different ages and gender interact with content production systems differently \cite{hu2007demographic,pandya2020use}. Therefore, it can be efficient to identify the users' demographics to provide content closer to their preferences. In this study, users' age and gender is modeled and predicted by using implicit methods. As shown in the results, identifying age by using $\alpha$-popular movies achieves acceptable accuracy against determining the user's age based on all ratings. In addition, a large amount of ratings are omitted for classification, which reduces the need of model updating over time. We also use these methods for gender prediction that show an improvement compared with other works.

Furthermore, in this study, it is shown that liked items provide appropriate information for modeling, and that the modeling accuracy obtained by using liked items is acceptable. Moreover, ratings related to liked items are only about a half of all ratings. Therefore, by training the models only based on the liked items, we eliminate almost 50\% of ratings.

Also, we show that users tend to register rates for popular movies in their early ratings, and after a while, this rate registration reduces over time. Because the built model needs an update when a user registers new ratings and updating these models is time-consuming, so, building demographic prediction models with popular movies reduces regular updating models.

In the future, to increase the accuracy of age prediction, other information related to popular items can be used. For instance, the comments that users make about popular items can be useful for identification improvement. We can also use the movies' synopsis to gain more information about the movies.


%
\section*{Conflict of interest}
The authors declare that there is no competing financial interests or personal relationships that influence the work in this paper .

\bibliographystyle{spmpsci}      
\bibliography{References}	

\begin{thebibliography}{10}
\providecommand{\url}[1]{{#1}}
\providecommand{\urlprefix}{URL }
\expandafter\ifx\csname urlstyle\endcsname\relax
  \providecommand{\doi}[1]{DOI~\discretionary{}{}{}#1}\else
  \providecommand{\doi}{DOI~\discretionary{}{}{}\begingroup
  \urlstyle{rm}\Url}\fi

\bibitem{noauthor_movielens_2013}
{MovieLens Dataset} (2013).
\newblock \urlprefix\url{https://grouplens.org/datasets/movielens/}

\bibitem{motion}
Film ratings (2020).
\newblock \urlprefix\url{https://www.motionpictures.org/film-ratings/}

\bibitem{imdb}
Imdb.
\newblock https://www.imdb.com/ (2020).
\newblock \urlprefix\url{https://www.imdb.com/}

\bibitem{imdbp}
Parental guide (2020).
\newblock
  \urlprefix\url{https://help.imdb.com/article/contribution/titles/parentalguide/GF4KYKYJA4PKQB32?ref\_=helpms\_helpart\_inline}

\bibitem{ahmadian2020social}
Ahmadian, S., Joorabloo, N., Jalili, M., Ren, Y., Meghdadi, M., Afsharchi, M.:
  A social recommender system based on reliable implicit relationships.
\newblock Knowledge-Based Systems \textbf{192}, 105371 (2020)

\bibitem{al2019predicting}
Al-Zuabi, I.M., Jafar, A., Aljoumaa, K.: Predicting customer’s gender and age
  depending on mobile phone data.
\newblock Journal of Big Data \textbf{6}(1), 1--16 (2019)

\bibitem{bin2017identifying}
Bin~Tareaf, R., Berger, P., Hennig, P., Jung, J., Meinel, C.: Identifying
  audience attributes: predicting age, gender and personality for enhanced
  article writing.
\newblock In: Proceedings of the 2017 International Conference on Cloud and Big
  Data Computing, pp. 79--88 (2017)

\bibitem{diez2019optimizing}
D{\'\i}ez, J., Mart{\'\i}nez-Rego, D., Alonso-Betanzos, A., Luaces, O.,
  Bahamonde, A.: Optimizing novelty and diversity in recommendations.
\newblock Progress in Artificial Intelligence \textbf{8}(1), 101--109 (2019)

\bibitem{dong2014inferring}
Dong, Y., Yang, Y., Tang, J., Yang, Y., Chawla, N.V.: Inferring user
  demographics and social strategies in mobile social networks.
\newblock In: Proceedings of the 20th ACM SIGKDD international conference on
  Knowledge discovery and data mining, pp. 15--24 (2014)

\bibitem{eirinaki2018recommender}
Eirinaki, M., Gao, J., Varlamis, I., Tserpes, K.: Recommender systems for
  large-scale social networks: A review of challenges and solutions (2018)

\bibitem{garcia2012prototype}
Garcia, S., Derrac, J., Cano, J., Herrera, F.: Prototype selection for nearest
  neighbor classification: Taxonomy and empirical study.
\newblock IEEE transactions on pattern analysis and machine intelligence
  \textbf{34}(3), 417--435 (2012)

\bibitem{garcia2020trend}
Garcia-Guzman, R., Andrade-Ambriz, Y.A., Ibarra-Manzano, M.A., Ledesma, S.,
  Gomez, J.C., Almanza-Ojeda, D.L.: Trend-based categories recommendations and
  age-gender prediction for pinterest and twitter users.
\newblock Applied Sciences \textbf{10}(17), 5957 (2020)

\bibitem{guimaraes2017age}
Guimaraes, R.G., Rosa, R.L., De~Gaetano, D., Rodriguez, D.Z., Bressan, G.: Age
  groups classification in social network using deep learning.
\newblock IEEE Access \textbf{5}, 10805--10816 (2017)

\bibitem{hamedani2019recommending}
Hamedani, E.M., Kaedi, M.: Recommending the long tail items through
  personalized diversification.
\newblock Knowledge-Based Systems \textbf{164}, 348--357 (2019)

\bibitem{hu2007demographic}
Hu, J., Zeng, H.J., Li, H., Niu, C., Chen, Z.: Demographic prediction based on
  user's browsing behavior.
\newblock In: Proceedings of the 16th international conference on World Wide
  Web, pp. 151--160 (2007)

\bibitem{huang2017age}
Huang, J., Li, B., Zhu, J., Chen, J.: Age classification with deep learning
  face representation.
\newblock Multimedia Tools and Applications \textbf{76}(19), 20231--20247
  (2017)

\bibitem{huang2019novel}
Huang, X., Wu, F.: A novel topic-based framework for recommending long tail
  products.
\newblock Computers \& Industrial Engineering \textbf{137}, 106063 (2019)

\bibitem{japkowicz2011evaluating}
Japkowicz, N., Shah, M.: Evaluating learning algorithms: a classification
  perspective.
\newblock Cambridge University Press (2011)

\bibitem{kalimeri2019predicting}
Kalimeri, K., Beir{\'o}, M.G., Delfino, M., Raleigh, R., Cattuto, C.:
  Predicting demographics, moral foundations, and human values from digital
  behaviours.
\newblock Computers in Human Behavior \textbf{92}, 428--445 (2019)

\bibitem{kim2019predicting}
Kim, I., Pant, G.: Predicting web site audience demographics using content and
  design cues.
\newblock Information \& Management \textbf{56}(5), 718--730 (2019)

\bibitem{li2019improving}
Li, Y., Yang, L., Xu, B., Wang, J., Lin, H.: Improving user attribute
  classification with text and social network attention.
\newblock Cognitive Computation \textbf{11}(4), 459--468 (2019)

\bibitem{malmi2016you}
Malmi, E., Weber, I.: You are what apps you use: Demographic prediction based
  on user's apps.
\newblock In: Proceedings of the International AAAI Conference on Web and
  Social Media, vol.~10 (2016)

\bibitem{morgan2017predicting}
Morgan-Lopez, A.A., Kim, A.E., Chew, R.F., Ruddle, P.: Predicting age groups of
  twitter users based on language and metadata features.
\newblock PloS one \textbf{12}(8), e0183537 (2017)

\bibitem{nguyen2013old}
Nguyen, D., Gravel, R., Trieschnigg, D., Meder, T.: " how old do you think i
  am?" a study of language and age in twitter.
\newblock In: Proceedings of the International AAAI Conference on Web and
  Social Media, vol.~7 (2013)

\bibitem{pandya2020use}
Pandya, A., Oussalah, M., Monachesi, P., Kostakos, P.: On the use of
  distributed semantics of tweet metadata for user age prediction.
\newblock Future Generation Computer Systems \textbf{102}, 437--452 (2020)

\bibitem{park2008long}
Park, Y.J., Tuzhilin, A.: The long tail of recommender systems and how to
  leverage it.
\newblock In: Proceedings of the 2008 ACM conference on Recommender systems,
  pp. 11--18 (2008)

\bibitem{sreepada2020mitigating}
Sreepada, R.S., Patra, B.K.: Mitigating long tail effect in recommendations
  using few shot learning technique.
\newblock Expert Systems with Applications \textbf{140}, 112887 (2020)

\bibitem{taeuscher2019uncertainty}
Taeuscher, K.: Uncertainty kills the long tail: demand concentration in
  peer-to-peer marketplaces.
\newblock Electronic Markets \textbf{29}(4), 649--660 (2019)

\bibitem{valcarce2016item}
Valcarce, D., Parapar, J., Barreiro, {\'A}.: Item-based relevance modelling of
  recommendations for getting rid of long tail products.
\newblock Knowledge-Based Systems \textbf{103}, 41--51 (2016)

\bibitem{wang2016multi}
Wang, S., Gong, M., Li, H., Yang, J.: Multi-objective optimization for long
  tail recommendation.
\newblock Knowledge-Based Systems \textbf{104}, 145--155 (2016)

\bibitem{zhong2013user}
Zhong, E., Tan, B., Mo, K., Yang, Q.: User demographics prediction based on
  mobile data.
\newblock Pervasive and mobile computing \textbf{9}(6), 823--837 (2013)

\end{thebibliography}

%
%

\end{document}